%% file: paper.tex
\documentclass[final]{new_tlp}

\usepackage{url}
\usepackage{fancyvrb}
\usepackage{verbatim}
\usepackage{graphicx}
\usepackage{amsmath}
\usepackage{proof-dashed}
\usepackage{pxfonts}

\input{fp-macros}

\newcommand\bcmdtab{\noindent\bgroup\tabcolsep=0pt%
  \begin{tabular}{@{}p{10pc}@{}p{20pc}@{}}}
\newcommand\ecmdtab{\end{tabular}\egroup}

\newcommand{\cmu}{\ensuremath{^\dag}}
\newcommand{\fcup}{\ensuremath{^\ddag}}

\title[Theory and Practice of Logic Programming]
{A Linear Logic Programming Language for Concurrent Programming over Graph Structures}

\author[Flavio Cruz, Ricardo Rocha, Seth Copen Goldstein and Frank Pfenning]
       {Flavio Cruz\cmu\fcup, Ricardo Rocha\fcup, Seth Copen Goldstein\cmu, Frank Pfenning\cmu\\
       \cmu Carnegie Mellon University, Pittsburgh, PA 15213\\
       \email{{fmfernan, seth, fp}@cs.cmu.edu} \\
       \fcup CRACS \& INESC TEC, Faculty of Sciences, University Of Porto\\
       Rua do Campo Alegre, 1021/1055, 4169-007 Porto, Portugal\\
       \email{ricroc@dcc.fc.up.pt}}

\submitted{13 February 2014}
\revised{18 April 2014}
\accepted{1 May 2014}
\jdate{January 2014}
\pubyear{2014}
\pagerange{\pageref{firstpage}--\pageref{lastpage}}

\newtheorem{lemma}{Lemma}[section]
\newtheorem{theorem}{Theorem}[section]

\begin{document}

\maketitle

\begin{abstract}
\input{abstract}
\end{abstract}



\section{Introduction}
\input{introduction}

\section{LM By Example}
\input{example}

\section{The LM Language}
\input{language}

\section{Some Sample LM Programs}
\input{programs}

\section{Proof Theory}
\input{proofs}

\section{Concurrency}
\input{concurrency}

\section{Related Work}
\input{related_work}

\section{Closing Remarks}
\input{conclusion}

\section*{Acknowledgments}
\input{ack}

\bibliographystyle{acmtrans}
\bibliography{refs}

\clearpage
\appendix
\input{appendix}

\end{document}

%% file: fp-macros.tex
\usepackage{latexsym}
\usepackage{amssymb}            
\usepackage{stmaryrd}           

\newcommand{\m}[1]{\mathsf{#1}}

\newcommand{\seq}{\vdash}

\newcommand{\lolli}{\multimap}

\newcommand{\with}{\mathbin{\binampersand}}

\newcommand{\one}{\mathbf{1}}

\newcommand{\bang}{{!}}


%% file: abstract.tex
We have designed a new logic programming language called LM (Linear Meld)
for programming graph-based algorithms in a declarative fashion. Our language
is based on linear logic, an expressive logical system where logical facts can
be consumed. Because LM integrates both classical and linear logic, LM tends
to be more expressive than other logic programming languages.
LM programs are naturally concurrent because facts are partitioned by nodes
of a graph data structure.
Computation is performed at the node level while communication happens
between connected nodes.
In this paper, we present the syntax and operational semantics of our language
and illustrate its use through a number of examples.

%% file: introduction.tex
Due to the popularity of social networks and the explosion of the content available in the World Wide Web, there has been
increased interest in running graph-based algorithms concurrently. Most of the available frameworks are implemented as libraries on top
of imperative programming languages, which require knowledge of both the library and the interface, making it difficult
for both novice and expert programmers to learn and use correctly.
Reasoning about the programs requires knowing how
the library schedules execution and the operational semantics of the underlying language.

Some good examples are the Dryad, Pregel and GraphLab systems.
The Dryad system~\cite{Isard:2007:DDD:1272996.1273005} is a framework that combines computational vertices
with communication channels (edges) to form a data-flow graph. Each program is scheduled to
run on multiple computers or cores and data is partitioned during runtime. Routines that run on computational vertices
are sequential, with no locking required.
The Pregel system~\cite{Malewicz:2010:PSL:1807167.1807184} is also graph-based, although programs have a more strict
structure. They must be represented as a sequence of iterations where each iteration is composed of computation and message passing.
Pregel is aimed at solving very big graphs
and to scale to large architectures. GraphLab~\cite{GraphLab2010} is a C++ library for developing parallel machine learning algorithms. While
Pregel uses message passing, GraphLab allows nodes to have read/write access to different scopes through different concurrent access models in order to balance performance and data consistency. Each consistency model provides different guarantees that are suited to multiple classes of algorithms. GraphLab also provides several schedulers that dictate the order in which node's are computed.

An alternative promising approach for graph-based algorithms 
is logic programming. For instance, the P2 system~\cite{Loo-condie-garofalakis-p2}, used Datalog to map nodes of a computer network
to a graph, where each node would do computation locally and could communicate with neighbor nodes.
Another good example is the Meld language, created by
Ashley-Rollman et al.~\cite{ashley-rollman-derosa-iros07wksp,ashley-rollman-iclp09}.
Meld was itself inspired in the P2 system but adapted to the concept of massively distributed systems made of modular robots with a
dynamic topology.
Logic-based systems are more amenable to proof since a program is just a set of logical clauses.

In this paper, we present a new logic programming language called LM (Linear Meld) for concurrent programming over graph structures designed to take advantage
of the recent architectures such as multicores or clusters of multicores. LM is based on the Meld language, but differs from other logic programming languages
such as Datalog or Prolog in three main aspects. First, it integrates both classical
logic and linear logic into the language, allowing some facts to be retracted and asserted in a logical fashion. Second, unlike
Prolog, LM is a bottom up logic programming language (similar to Datalog) since the database is updated incrementally as rules are
applied. Third, LM is a language created to solve general graph-based algorithms, unlike P2 or Meld which were designed for more specific domains.

In the following sections, we present the syntax and semantics of our language and explain how to write programs that take advantage of its expressive power. We identify three key contributions in our work:

\begin{description}
   \item[Linear Logic:] We integrate linear logic into the original Meld language so that program state can be encoded naturally.
   Meld started as a classical logic programming language where everything that is derived is true until the end
   of the execution. Linear logic turns logical facts into resources that will be consumed when a rule is applied. In turn, this makes it possible to represent program state in a natural and declarative fashion.
   \item[Concurrency:] LM programs are naturally concurrent because facts are partitioned by vertices of a graph data structure. While the original Meld sees graphs as a network of robots, we see each node as a member of a distributed data structure. This is made possible due to the
   restrictions on derivation rules which only use local facts but also permit node communication.
   \item[Semantics:] Starting from a fragment of linear logic used in LM, we formalize a high level dynamic semantics that is closely related to this fragment.
   We then design a low level dynamic semantics and sketch the soundness proof of our low level
   semantics with respect to the high level language specification. The low level specification
   provides the basis for a correct implementation of LM.
\end{description}

To realize LM, we have implemented a compiler and a virtual machine that executes LM programs on multicore machines
\footnote{Source code is available at \url{http://github.com/flavioc/meld}.}. We also have a preliminary version that runs on networks by
using OpenMPI as a communication layer. Our experimental results show that LM has good scalability.
Several interesting programs were implemented such as belief propagation~\cite{Gonzalez+al:aistats09paraml},
belief propagation with residual splash~\cite{Gonzalez+al:aistats09paraml}, PageRank, graph coloring,
N queens, shortest path, diameter estimation, map reduce, game of life, quick-sort, neural network training, among others.
While these results are evidence that LM is a promising language, this paper will only focus on the more formal aspects of our work.

%% file: example.tex
Linear Meld (LM) is a \emph{forward chaining} logic programming language in the style of Datalog~\cite{Ullman:1990:PDK:533142}. The program is defined as a \emph{database of facts} and a set of \emph{derivation rules}.
Initially, we populate the database with the program's axioms and then determine which derivation rules can be applied by using the current database. Once a rule is applied, we derive new facts, which are then added to the database.
If a rule uses linear facts, they are consumed and thus deleted from the database.
The program stops when we reach \emph{quiescence}, that is, when we can no longer
apply any derivation rule.

The database of facts can be seen as a graph data structure where each node or vertex contains a
fraction of the database.  Since derivation rules can only manipulate facts belonging to
a node, we are able to perform independent rule derivations.

Each fact is a predicate on a tuple of \emph{values}, where the type of the predicate prescribes the types of the arguments.
LM rules are type-checked using the predicate declarations in the header of the program. LM has a simple type system that includes types such as
\emph{node}, \emph{int}, \emph{float}, \emph{string}, \emph{bool}. Recursive types such as \emph{list X} and \emph{pair X; Y} are
also allowed.

The first argument of every predicate must be typed as a \emph{node}.
For concurrency and data partitioning purposes, derivation rules are constrained by the expressions that can be written in the body.
The body of every rule can only refer to facts in the same node (same first argument).
However, the expressions in the head may refer to other nodes, as long as those nodes are instantiated in the body of the rule.

Each rule in LM has a defined priority that is inferred from its position in the source file.
Rules at the beginning of the file have higher priority. At the node level, we consider all
the new facts that have been not consider yet to create a set of \emph{candidate rules}.
The set of candidate rules is then applied (by priority) and updated as new facts are derived.

Our first program example is shown in Fig.~\ref{code:message}. This is a message routing program
that simulates message transmission through a network of nodes.
We first declare all the predicates (lines 1-2), which represent the different facts we are going to
use.
Predicate \texttt{edge/2} is a non \texttt{linear} (persistent) predicate and \texttt{message/3} is linear. While linear facts may be retracted, persistent facts are always
true once they are derived.

The program rules are declared in lines 4-8, while the program's axioms are written in lines 10-11.
The general form of a rule is $\mathtt{A_1},...,\mathtt{A_n}$ \texttt{-o} $\mathtt{B_1},...,\mathtt{B_m}$, where $\mathtt{A_1},...,\mathtt{A_n}$ are matched against local facts and $\mathtt{B_1},...,\mathtt{B_m}$ are locally asserted or transmitted to a neighboring node.
When persistent facts are used (line 4) they must be
preceded by \texttt{!} for readability.

\begin{figure}[h!]
\scriptsize\begin{Verbatim}[numbers=left]
type edge(node, node). // define direct edge
type linear message(node, string, list node). // message format

message(A, Content, [B | L]), !edge(A, B)
   -o message(B, Content, L). // message derived at node B

message(A, Content, [])
   -o 1. // message received

!edge(@1, @2). !edge(@2, @3). !edge(@3, @4). !edge(@1, @3).
message(@1, 'Hello World', [@3, @4]).
\end{Verbatim}
\caption{Message program.}
  \label{code:message}
\end{figure}
\normalsize

The first rule (lines 4-5) grabs the next node in the route list (third argument of \texttt{message/3}),
ensures that a communication edge exists with \texttt{!edge(A, B)} and
derives a new \texttt{message(B,~Content,~L)} fact at node \texttt{B}.
When the route list is empty, the message has reached its destination and thus it is consumed
(rule in lines 7-8). Note that the '\texttt{1}' in the head of the rule on line 8 means that nothing is derived.

Figure~\ref{code:visit} presents another complete LM program which given a graph
of nodes visits all nodes reachable from node $@1$.
The first rule of the program (lines 6-7) is fired when a node \texttt{A} has both the \texttt{visit(A)} and \texttt{unvisited(A)} facts.
When fired, we first derive \texttt{visited(A)} to mark node \texttt{A} as \textit{visited} and use a
\emph{comprehension} to go through all the edge facts \texttt{edge(A,B)} and derive \texttt{visit(B)} for each
one (comprehensions are explained next in detail). This forces those nodes to be visited.
The second rule (lines 9-10) is fired when a
node \texttt{A} is already visited more than once: we keep the \texttt{visited(A)} fact and delete \texttt{visit(A)}.
Line 14 starts the process by asserting the \texttt{visit(@1)} fact.

\begin{figure}[h!]
\scriptsize\begin{Verbatim}[numbers=left]
type edge(node, node).
type linear visit(node).
type linear unvisited(node).
type linear visited(node).

visit(A), unvisited(A)
   -o visited(A), {B | !edge(A, B) | visit(B)}. // mark node as visited and visit neighbors

visit(A), visited(A)
   -o visited(A). // already visited

!edge(@1, @2). !edge(@2, @3). !edge(@1, @4). !edge(@2, @4).
unvisited(@1). unvisited(@2). unvisited(@3). unvisited(@4).
visit(@1).
\end{Verbatim}
  \caption{Visit program.}
  \label{code:visit}
\end{figure}
\normalsize

If the graph is connected, it is easy to prove that every node \texttt{A} will derive \texttt{visited(A)},
regardless of the order in which rules are applied.

%% file: language.tex
\newcommand{\selector}[0]{[\; S \Rightarrow y; \; BE \;] \lolli HE}
\newcommand{\comprehension}[0]{\{ \; \widehat{x}; \; BE; \; SH \; \}}
\newcommand{\aggregate}[0]{[\; A \Rightarrow y; \; \widehat{x}; \; BE; \; SH_1; \; SH_2 \;]}

Table~\ref{tbl:ast} shows the abstract syntax for rules in LM.
An LM program $Prog$ consists of a set of derivation rules $\Sigma$ and a database $D$.
A derivation rule $R$ may be written as $BE \lolli HE$ where $BE$ is the body of the rule and
$HE$ is the head.
We can also explicitly universally quantify over variables in a rule using $\; \forall_{x}. R$.
If we want to control how facts are selected in the body, we may use \emph{selectors} of
the form $\selector$ (explained later).

\begin{table}[h]
\centering
\begin{tabular}{ l l c l }
  Program & $Prog$ & $::=$ & $\Sigma, D$ \\
  Set Of Rules & $\Sigma$ & $::=$ & $\cdot \; | \; \Sigma, R$\\
  Database & $D$ & $::=$ & $\Gamma; \Delta$ \\
  Rule & $R$ & $::=$ & $BE \lolli HE \; | \; \forall_{x}. R \; | \; \selector$ \\
  Body Expression & $BE$ & $::=$ & $L \; | \; P \; | \; C \; | \; BE, BE \; | \; \exists_{x}. BE \; | \; 1$\\
  Head Expression & $HE$ & $::=$ & $L \; | \; P \; | \; HE, HE \; | \; EE \; | \; CE \; | \; AE \; | \; 1$\\
  
  Linear Fact & $L$ & $::=$ & $l(\hat{x})$\\
  Persistent Fact & $P$ & $::=$ & $\bang p(\hat{x})$\\
  Constraint & $C$ & $::=$ & $c(\hat{x})$ \\
  Selector Operation & $S$ & $::=$ & $\mathtt{min} \; | \; \mathtt{max} \; | \; \mathtt{random}$\\
  
  Exists Expression & $EE$ & $::=$ & $\exists_{\widehat{x}}. SH$ \\
  Comprehension & $CE$ & $::=$ & $\comprehension$ \\
  Aggregate & $AE$ & $::=$ & $\aggregate$ \\
  Aggregate Operation & $A$ & $::=$ & $\mathtt{min} \; | \; \mathtt{max} \; | \; \mathtt{sum} \; | \; \mathtt{count}$ \\
  
  Sub-Head & $SH$ & $::=$ & $L \; | \; P \; | \; SH, SH \; | \; 1$\\
  
  Known Linear Facts & $\Delta$ & $::=$ & $\cdot \; | \; \Delta, l(\hat{t})$ \\
  Known Persistent Facts & $\Gamma$ & $::=$ & $\cdot \; | \; \Gamma, \bang p(\hat{t})$ \\
\end{tabular}
\caption{Abstract syntax of LM.}\label{tbl:ast}
\end{table}

The body of the rule, $BE$, may contain linear ($L$) and persistent ($P$) \emph{fact expressions} and
constraints ($C$). We can chain those elements by using $BE, BE$ or introduce body variables using $\exists_{x}. BE$.
Alternatively we can use an empty body by using $1$, which creates an axiom.

Fact expressions are template facts that instantiate variables
(from facts in the database) such as \texttt{visit(A)} in line 10 in Fig.~\ref{code:visit}.
Constraints are boolean expressions that must
be true in order for the rule to be fired (for example, \texttt{C~=~A~+~B}). Constraints use variables from fact expressions and are built using a small functional language that includes mathematical operations, boolean operations, external functions and literal values.

The head of a rule ($HE$) contains linear ($L$) and persistent ($P$) \emph{fact templates} which are uninstantiated facts and will derive new facts. The head can also have \emph{exist expressions} ($EE$), \emph{comprehensions} ($CE$) and \emph{aggregates} ($AE$). All those expressions
may use all the variables instantiated in the body. We can also use an empty head by choosing $1$.

\paragraph{Selectors}

When a rule body is instantiated using facts from the database, facts are picked
non-deterministically. While our system uses an implementation dependent order for
efficiency reasons, sometimes it is important to sort facts by one of the arguments
because linearity imposes commitment during rule derivation. The abstract syntax for
this expression is $\selector$, where $S$ is the selection operation and $y$ is the
variable in the body $BE$ that represents the value to be selected according to $S$.
An example using concrete syntax is as follows:

{\footnotesize
\begin{Verbatim}
[min => W | !edge(A, B), weight(A, B, W)] -o picked(A, B, W).
\end{Verbatim}
}

In this case, we order the \texttt{weight} facts by \texttt{W} in ascending order and then try
to match them. Other operations available are \texttt{max} and \texttt{random} (to force no pre-defined order).

\paragraph{Exists Expression}

Exists expressions ($EE$) are based on the linear logic term of the same name and are used to create new node addresses.
We can then use the new address to instantiate new facts for this node.  
The following example illustrates the use of the exists expression, where we derive
\texttt{perform-work} at a new node \texttt{B}.

{\footnotesize
\begin{Verbatim}
do-work(A, W) -o exists B. (perform-work(B, W)).
\end{Verbatim}
}

\paragraph{Comprehensions}

Sometimes we need to consume a linear fact and then immediately generate several facts depending on
the contents of the database. To solve this particular need, we created the concept of comprehensions, which are
sub-rules that are applied with all possible combinations of facts from the database. In a comprehension $\comprehension$, $\widehat{x}$ is a list of variables, $BE$ is the comprehension's body and $SH$ is the head.
The body $BE$ is used to generate all possible combinations for the head $SH$, according to the facts
in the database. Note that $BE$ is also locally restricted.

We have already seen an example of comprehensions in the visit program (Fig.~\ref{code:visit} line 7).
Here, we match \texttt{!edge(A, B)} using all the combinations available in the database and derive \texttt{visit(B)}
for each combination.

\paragraph{Aggregates}

Another useful feature in logic programs is the ability to reduce several facts into a single fact.
In LM we have aggregates ($AE$), a special kind of sub-rule similar to comprehensions.
In the abstract syntax $\aggregate$, $A$ is the aggregate operation, $\widehat{x}$ is the list of variables
introduced in $BE$, $SH_1$ and $SH_2$ and $y$ is the variable in the body
$BE$ that represents the values to be aggregated using $A$.
We use $\widehat{x}$ to try all the combinations of $BE$, but, in addition to deriving $SH_1$ for each combination,
we aggregate the values represented by $y$ and derive $SH_2$ only once using $y$.

Let's consider a database with the following facts and a rule:

{\footnotesize
\begin{Verbatim}
price(@1, 3). price(@1, 4). price(@1, 5).
count-prices(@1).
count-prices(A) -o [sum => P | . | price(A, P) | 1 | total(A, P)].
\end{Verbatim}
}

By applying the rule, we consume \texttt{count-prices(@1)} and
derive the aggregate which consumes all the \texttt{price(@1, P)} facts.
These are added and \texttt{total(@1,~12)} is derived.
LM provides aggregate operations such as \texttt{min} (minimum), \texttt{max} (maximum), \texttt{sum} and \texttt{count}.

%% file: programs.tex
We now present LM programs in order to illustrate common programming techniques\footnote{More examples of LM programs are available at \url{http://github.com/flavioc/meld}.}.

\paragraph{Shortest Distance}

Finding the shortest distance between two nodes in a graph is another well known graph problem.
Fig.~\ref{code:shortest_path} presents the LM code to solve this particular problem.

We use an \texttt{edge/3}
predicate to represent directed edges between nodes and their corresponding weights. To represent the shortest
distance to a node \texttt{startnode} we have a \texttt{path(A,D,F)} where \texttt{D} is the distance to \texttt{startnode}
and \texttt{F} is a flag to indicate if such distance has been propagated to the neighbors. Since the distance from
the \texttt{startnode} to itself is \texttt{0}, we start the algorithm with the axiom \texttt{path(startnode,0,notused)}.

The first rule avoids propagating paths with the same distance and the second rule eliminates paths where the distance
is already larger than some other distance. Finally, the third rule, marks the path as \texttt{used} and propagates
the distance to the neighboring nodes by taking into account the edge weights.
Eventually, the program will reach quiescence and the shortest distance between \texttt{startnode} and \texttt{finalnode}
will be determined.

\newcommand{\BigO}[1]{\ensuremath{\operatorname{O}\bigl(#1\bigr)}}

\begin{figure}[h]
\scriptsize\begin{Verbatim}[numbers=left]
type edge(node, node, int).
type linear path(node, int, int).

const used = 1.
const notused = 0.

path(startnode, 0, notused).

path(A, D, used), path(A, D, notused)
   -o path(A, D, used).

path(A, D1, X), path(A, D2, Y), D1 <= D2
   -o path(A, D1, X). // keep the shorter distance

path(A, D, notused), A <> finalnode
   -o {B, W | !edge(A, B, W) | path(B, D + W, notused)}, path(A, D, used). // propagate new distance
\end{Verbatim}
\caption{Shortest Distance Program.}
\label{code:shortest_path}
\end{figure}
\normalsize

In the worst case, this algorithm runs in \BigO{N E}, where $N$ is the number of nodes and $E$ is the
number of edges. If we decide to always propagate the shortest distance of the graph, we get Dijkstra's algorithm~\cite{Dijkstra}. However, this is
not feasible, since we would need to globally decide which node to run next, removing concurrency.

\paragraph{PageRank}

PageRank~\cite{Page:2001:MNR} is a well known graph algorithm that is used to compute the relative relevance of web pages.
The code for a synchronous version of the algorithm is shown in Fig.~\ref{code:pagerank}.
As the name indicates, the pagerank is computed for a certain number of iterations. The initial pagerank is the same for every page and is
initialized in the first rule (line 12) along with an accumulator.

\begin{figure}[h]
   \scriptsize\begin{Verbatim}[numbers=left]
type output(node, node, float).
type linear pagerank(node, float, int).
type numLinks(node, int).
type numInput(node, int).
type linear accumulator(node, float, int, int).
type linear newrank(node, node, float, int).
type linear start(node).

start(A).

start(A), !numInput(A, T)
   -o accumulator(A, 0.0, T, 1), pagerank(A, 1.0 / float(@world), 0).

pagerank(A, V, Id), !numLinks(A, C), Id < iterations, Result = V / float(C)
   -o {B, W | !output(A, B, W) | newrank(B, A, Result, Id + 1)}. // propagate new pagerank value

accumulator(A, Acc, 0, Id), !numInput(A, T), V = 0.85 + 0.15 * Acc, Id <= iterations
   -o pagerank(A, V, Id), accumulator(A, 0.0, T, Id + 1). // new pagerank value
	
newrank(A, B, V, Id), accumulator(A, Acc, T, Id), T > 0
   -o [sum => S, count => C | D | newrank(A, D, S, Id) | 1 | accumulator(A, Acc + V + S, T - 1 - C, Id)].
\end{Verbatim}
\caption{Synchronous PageRank program.}
\label{code:pagerank}
\normalsize
\end{figure}

The second rule of the program (lines 14-15) propagates a newly computed pagerank value to all neighbors. Each node will then accumulate
the pagerank values that are sent to them through the fourth rule (lines 20-21) and it will immediately add other currently available values
through the use of the aggregate. When we have accumulated all the values we need, the third rule (lines 17-18) is fired and a new pagerank value is derived.


\paragraph{N-Queens}

The N-Queens~\cite{8queens} puzzle is the problem of placing N chess queens on an NxN chessboard so
that no pair of two queens attack each other. The specific challenge of finding all the distinct
solutions to this problem is a good benchmark in designing parallel algorithms. The LM solution is presented
in \ref{code:nqueens}.

First, we consider each cell of the chessboard as a node that can communicate with the adjacent left
(\texttt{left}) and adjacent right (\texttt{right}) cells and also with the first two non-diagonal cells in the next row
(\texttt{down-left} and \texttt{down-right}). For instance, the node at cell \texttt{(0,~3)} (fourth cell in the first row) will connect
to cells \texttt{(0,~2)}, \texttt{(0,~4)} and also \texttt{(1,~1)} and \texttt{(1,~5)}, respectively. The states are represented as a list
of integers, where each integer is the column number where the queen was placed. For example \texttt{[2, 0]}
means that a queen is placed in cell \texttt{(0,~0)} and another in cell \texttt{(1,~2)}.

An empty state is instantiated in the top-left node \texttt{(0,~0)} and then propagated to all nodes in the same row (lines 19-20).
Each node then tries to place a queen on their cell and then send a new state to the row below (lines 52-54).
Recursively, when a node receives a new state, it will (i) send the state to the left
or to the right and (ii) try to place the queen in its cell (using \texttt{test-y}, \texttt{test-diag-left} and \texttt{test-diag-right}). When a cell cannot place a queen, that state is deleted (lines 29, 37 and 45).
When the program ends, the states will be placed in the bottom row (lines 49-50).

Most parallel implementations distribute the search space of the problem by assigning incomplete boards as tasks to workers.
Our approach is unusual because our tasks are the cells of the board.

%% file: proofs.tex
\newcommand{\mz}{\m{match} \;}
\newcommand{\tab}[0]{\;\;\;\;}
\newcommand{\dz}{\m{derive} \;}
\newcommand{\comp}[0]{\m{comp} \;}
\newcommand{\az}{\m{apply} \;}
\newcommand{\doz}{\m{run} \;}
\newcommand{\seqnocut}[3]{#1 ; #2 \Rightarrow #3}
\newcommand{\defeq}{\buildrel\triangle\over =}
\newcommand{\compr}[1]{\m{def} \; #1}

\newcommand{\mo}{\m{match}_{LLD} \;}
\newcommand{\cont}{\m{cont}_{LLD} \;}
\newcommand{\contc}{\m{cont}_{LLDc} \;}
\newcommand{\done}{\m{derive}_{LLD} \;}
\newcommand{\doo}{\m{run}_{LLD} \;}
\newcommand{\mc}[0]{\m{match}_{LLDc} \; }
\newcommand{\dall}[0]{\m{fix}_{LLD} \; }
\newcommand{\strans}[0]{\m{update}_{LLD} \;}
\newcommand{\dc}{\m{derive}_{LLDc} \;}
\newcommand{\ao}{\m{apply}_{LLD} \;}

We now present the sequent calculus of a fragment of intuitionistic linear
logic~\cite{girard-87} used by LM followed by the dynamic semantics of LM built on top of
this fragment.

We use a standard set of connectives except the $\m{def} \; A$ connective, which is inspired on Baelde's work on least and
greatest fixed points in linear logic~\cite{Baelde:2012:LGF:2071368.2071370} and is used to logically justify
comprehensions and aggregates.
The sequent calculus (\ref{linear_logic}) has the form $\Psi; \Gamma; \Delta \rightarrow C$, where $\Psi$ is the typed term context used in the
quantifiers, $\Gamma$ is the set of persistent terms, $\Delta$ is the multi-set of linear propositions and $C$ is the proposition
to prove. Table~\ref{table:linear} relates linear logic with LM.

\begin{center}
\begin{table}[h]\resizebox{12cm}{!}{
    \begin{tabular}{ c l c l}
    \hline
    Connective                   & Description                                        & LM Place   & LM Example                           \\ \hline \hline
    $\emph{fact}(\hat{x})$       & Linear facts.                                      & Body/Head  & \texttt{path(A, P)}                  \\ 
    $\bang \emph{fact}(\hat{x})$ & Persistent facts.                                  & Body/Head  & \texttt{$\bang$edge(X, Y, W)}        \\ 
    $1$                          & Represents rules with an empty head.               & Head       & \texttt{1}                           \\ 
    $A \otimes B$                & Connect two expressions.                           & Body/Head  & \texttt{p(A), e(A, B)}               \\ 
    $\forall x. A$               & For variables defined in the rule.                 & Rule       & \texttt{p(A) $\lolli$ r(A)}          \\ 
    $\exists x. A$               & Instantiates new node variables.                   & Head       & \texttt{exists A.(p(A, P))}          \\ 
    $A \lolli B$                 & $\lolli$ means "linearly implies".                 & Rule       & \texttt{p(A, B) $\lolli$ r(A, B)}    \\ 
                                 & $A$ is the body and $B$ is the head.               &            &                                      \\ 
    $\m{def} A. B$               & Extension called definitions. Used for             & Head       & \texttt{\{B | !e(A, B) | v(B)\}}     \\ 
                                 & comprehensions and aggregates.                     &            &                                      \\ \hline
    \end{tabular}}
\caption{Connectives from Linear Logic used in LM.}
\label{table:linear}
\end{table}
\end{center}

In a comprehension, we want to apply an implication to as many matches as the database allows.
Our approach is to use definitions: given a comprehension $C = \{ \; \widehat{x}; \; A; \; B \; \}$ with a body $A$ and a head $B$, then we can build the following recursive definition:

\[
\compr{C} \defeq 1 \with ((A \lolli B) \otimes \compr{C})
\]

We unfold $\compr{C}$ to either stop (by selecting $1$) or get a linear implication $A \lolli B$
and a recursive definition. This uses linear logic's additive conjunction $\with$.
This form of definition does not capture the desired \emph{maximality} aspect of the comprehension,
since it commits to finding a particular form of proof and not all possible proofs. The low level
operational semantics will ensure maximality.

Aggregates work identically, but they need an extra argument to accumulate the aggregated value. If a sum aggregate $C$ has the form $[\;\m{sum} \Rightarrow y; \; \widehat{x}; \; A; \; B_1; \; B_2 \;]$, then the definition will be as follows (the aggregate is initiated as $\compr{C} \; 0$):

\[
\compr{C} \; V \defeq (\lambda v. B_2)V \with (\forall x. ((Ax \lolli B_1) \otimes \compr{C} \; (x + V)))
\]

\paragraph{Dynamic Semantics}

The dynamic semantics formalize the mechanism of matching and deriving a single rule at the node level. The semantics receive the node database and the program's rules as inputs and return as outputs the consumed linear facts, derived linear facts and derived persistent facts. Then, it is possible
to compute the program as a sequence of steps, by updating the database through sending or asserting.

\paragraph{High Level Dynamic Semantics}

The High Level Dynamic~(HLD) Semantics are closely related to the linear logic fragment presented above.
From the sequent calculus, we consider $\Gamma$ and $\Delta$ the database of persistent and linear facts, respectively.
We consider the rules of the program as persistent linear implications of the form $\bang (A \lolli B)$ that we put in a separate context $\Phi$.
We ignore the right hand side $C$ of the calculus and use inversion on the $\Delta$ and $\Gamma$ contexts so that we only have atomic terms (facts). To apply rules
we use chaining by focusing~\cite{Andreoli92logicprogramming} on the derivation rules of $\Phi$.
The HLD semantics are shown in Fig.~\ref{hld_semantics} and are composed of four judgments:

\begin{figure}[h]
   {\scriptsize
\[
\infer[\doz rule]
{\doz \Gamma ; \Delta ; R, \Phi \rightarrow \Xi' ; \Delta' ; \Gamma'}
{\az \Gamma ; \Delta ; R \rightarrow \Xi' ; \Delta' ; \Gamma'}
\tab
\infer[\az rule]
{\az \Gamma ; \Delta, \Delta'' ; A \lolli B \rightarrow \Xi' ; \Delta' ; \Gamma'}
{\mz \Gamma ; \Delta \rightarrow A & \dz \Gamma ; \Delta''; \Delta; \cdot ; \cdot ; B \rightarrow \Xi' ; \Delta' ; \Gamma'}
\]

\[
\infer[\mz 1]
{\mz \Gamma; \cdot \rightarrow 1}
{}
\tab
\infer[\mz p]
{\mz \Gamma; p \rightarrow p }
{}
\tab
\infer[\mz \bang p]
{\mz \Gamma, p; \cdot \rightarrow \bang p}
{}
\]

\[
\infer[\mz \otimes]
{\mz \Gamma; \Delta_1, \Delta_2 \rightarrow A \otimes B}
{\mz \Gamma; \Delta_1 \rightarrow A & \mz \Delta_2 \rightarrow B}
\]

\[
\infer[\dz p]
{\dz \Gamma ; \Delta ; \Xi ; \Gamma_1 ; \Delta_1 ; p, \Omega \rightarrow \Xi' ; \Delta' ; \Gamma'}
{\dz \Gamma ; \Delta ; \Xi ; \Gamma_1 ; p, \Delta_1 ; \Omega \rightarrow \Xi' ; \Delta' ; \Gamma'}
\tab
\infer[\dz \bang p]
{\dz \Gamma ; \Delta ; \Xi ; \Gamma_1 ; \Delta_1 ; \bang p, \Omega \rightarrow \Xi' ; \Delta' ; \Gamma'}
{\dz \Gamma ; \Delta ; \Xi ; \Gamma_1, p ; \Delta_1 ; \Omega \rightarrow \Xi' ; \Delta' ; \Gamma'}
\]

\[
\infer[\dz \otimes]
{\dz \Gamma ; \Delta ; \Xi ; \Gamma_1 ; \Delta_1 ; A \otimes B, \Omega \rightarrow \Xi' ; \Delta' ; \Gamma'}
{\dz \Gamma ; \Delta ; \Xi ; \Gamma_1 ; \Delta_1 ; A, B, \Omega \rightarrow \Xi' ; \Delta' ; \Gamma'}
\tab
\infer[\dz 1]
{\dz \Gamma ; \Delta ; \Xi ; \Gamma_1; \Delta_1 ; 1, \Omega \rightarrow \Xi' ; \Delta' ; \Gamma'}
{\dz \Gamma ; \Delta ; \Xi ; \Gamma_1; \Delta_1 ; \Omega \rightarrow \Xi' ; \Delta' ; \Gamma'}
\]

\[
\infer[\dz end]
{\dz \Gamma ; \Delta ; \Xi' ; \Gamma' ; \Delta' ; \cdot \rightarrow \Xi' ; \Delta' ; \Gamma'}
{}
\tab
\infer[\dz \lolli]
{\dz \Gamma ; \Delta_a, \Delta_b ; \Xi ; \Gamma_1 ; \Delta_1 ; A \lolli B, \Omega \rightarrow \Xi' ; \Delta' ; \Gamma'}
{\mz \Gamma ; \Delta_a \rightarrow A & \dz \Gamma ; \Delta_b ; \Xi, \Delta_a ; \Gamma_1 ; \Delta_1 ; B, \Omega \rightarrow \Xi' ; \Delta' ; \Gamma'}
\]

\[
\infer[\dz comp]
{\dz \Gamma ; \Delta ; \Xi ; \Gamma_1 ; \Delta_1 ; \comp A \lolli B, \Omega \rightarrow \Xi' ; \Delta' ; \Gamma'}
{\dz \Gamma ; \Delta ; \Xi ; \Gamma_1 ; \Delta_1 ; 1 \with (A \lolli B \otimes \comp A \lolli B), \Omega \rightarrow \Xi' ; \Delta' ; \Gamma'}
\]

\[
\infer[\m{derive} \with L]
{\dz \Gamma ; \Delta ; \Xi ; \Gamma_1 ; \Delta_1 ; A \with B, \Omega \rightarrow \Xi' ; \Delta'; \Gamma'}
{\dz \Gamma ; \Delta ; \Xi ; \Gamma_1 ; \Delta_1 ; A, \Omega \rightarrow \Xi' ; \Delta'; \Gamma'}
\tab
\infer[\m{derive} \with R]
{\dz \Gamma ; \Delta ; \Xi ; \Gamma_1 ; \Delta_1 ; A \with B, \Omega \rightarrow \Xi' ; \Delta' ; \Gamma'}
{\dz \Gamma ; \Delta ; \Xi ; \Gamma_1 ; \Delta_1 ; B, \Omega \rightarrow \Xi' ; \Delta' ; \Gamma'}
\]
}
\caption{High Level Dynamic Semantics.}
\label{hld_semantics}
\end{figure}

\begin{enumerate}
   \item $\doz \Gamma; \Delta; \Phi \rightarrow \Xi'; \Delta'; \Gamma'$ picks a rule from $\Phi$ and applies it using facts from $\Gamma$ and $\Delta$. $\Xi'$, $\Delta'$ and $\Gamma'$ are the outputs of the derivation process. $\Xi'$ are the linear facts consumed, $\Delta'$ are the linear facts derived and $\Gamma'$ the new persistent facts;
   \item $\az \Gamma; \Delta; R \rightarrow \Xi'; \Delta'; \Gamma'$ picks a subset of linear facts from $\Delta$ and matches the body of the rule $R$ and then derives the head;
   \item $\mz \Gamma; \Delta \rightarrow A$ verifies that all facts in $\Delta$ (set of consumed linear facts) prove $A$, the body of the rule. The context $\Gamma$ will be used to prove any persistent term in $A$;
   \item $\dz \Gamma; \Delta; \Xi; \Gamma_1; \Delta_1, \Omega \rightarrow \Xi'; \Delta'; \Gamma'$ deconstructs and instantiates the ordered head terms $\Omega$ (we start with the head of the rule $B$) and adds them to $\Delta_1$ and $\Gamma_1$, the contexts for the newly derived linear and persistent facts, respectively.
\end{enumerate}

Comprehensions are derived by non-deterministically deciding to apply the comprehension ($\m{derive} \with L$ and $\m{derive} \with R$)
and then using the $\mz$judgment in the rule $\dz comp$. We note that the HLD semantics do not take distribution into account,
since we assume that the database is global. We do not deal with unification or quantifiers since this is a well understood problem~\cite{Baader:1994:UT:185705.185711}.

\paragraph{Low Level Dynamic Semantics} The Low Level Dynamic~(LLD) Semantics shown in \ref{low_level_semantics} improve upon HLD by adding rule priorities, by removing non-determinism
when matching the body of rules by modeling all the matching steps and by applying comprehensions or aggregates as many times as the database allows.
Selectors can also be trivially implemented in LLD, although they are not shown in paper.

In LLD we try all the rules in order. For each rule, we use a \emph{continuation stack} to store the \emph{continuation frames} created by
each fact template $p$ present
in the body of the rule. Each frame considers all the facts relevant to the template given the current variable bindings ($\mo$rules), that
may or not fail during the remaining matching process. If we fail, we backtrack to try other alternatives (through $\cont$rules). If the
continuation stack becomes empty, we backtrack to try the next rule (rule $\cont next \; rule$). When we succeed the facts consumed are known
($\mo end$).

The derivation process in LLD is similar to the one used in HDL, except for the case of comprehensions or aggregates. For such cases
($\done comp$), we need to create a continuation stack and start matching the body of the expression as we did before.
When we match the body ($\mc$judgment), we fully derive the head ($\dc$judgment) and then we reuse the continuation stack to find which
other combinations of the database facts can be consumed ($\dc end$). By definition, the continuation stack contains
enough information to go through all combinations in the database.

However, in order to reuse the stack, we need to \emph{fix} it by removing all the frames pushed after the first continuation frame
of a linear fact. If we tried to use those frames, we would assumed that the linear facts used by the other frames were still in the database, but that
is not true because they have been consumed during the first application of the comprehension.
For example, if the body is $\bang\mathtt{a(X), b(X), c(X)}$ and the continuation stack has three frames (one per fact), we cannot backtrack to the frame of $\mathtt{c(X)}$
since at that point the matching process was assuming that the previous \texttt{b(X)} linear fact was still available.
Moreover, we also remove the consumed linear facts
from the frames of \texttt{b(X)} and $\bang$\texttt{a(X)} in order to make the stack fully consistent with the new database.
This is performed by rules using the $\strans$and $\dall$judgments.

We finally stop applying the comprehension when the continuation stack is empty ($\contc end$). 
Aggregates use the same mechanism as comprehensions, however we also need to keep track of the accumulated value.

\paragraph{Soundness}

The soundness theorem proves that if a rule was successfully derived in the LLD semantics then it can also be derived in the HLD semantics. The completeness theorem cannot
be proven since LLD lacks the non-determinism inherent in HLD.

We need prove to prove matching and derivation soundness of LLD in relation to HLD. The matching soundness lemma uses induction on the size of the continuation frames, the size of the continuation stack and the size of terms to match.

The derivation soundness lemma is trivial except for the case of comprehensions and aggregates. For such cases we use a modified version of the matching soundness theorem applied to the comprehension's body. It gives us $n$ $\mz$and $n$ $\dz$proofs (for maximality) that are used to rebuild the full derivation proof in HLD. This theorem is proved by
induction on the size of the continuation stack and continuation frames and uses lemmas that prove the correctness of the continuation stack after each application.\footnote{Details can be
found in \url{https://github.com/flavioc/formal-meld/blob/master/doc.pdf?raw=true}.}

%% file: concurrency.tex
Due to the restrictions on LM rules and the partitioning of facts across the graph, nodes are able to
run rules independently without using other node's facts. Node computation follows a \emph{don't care} or \emph{committed choice} non-determinism
since any node can be picked to run as long as it contains enough facts to fire a derivation rule.
Facts coming from other nodes will arrive in order of derivation but may be considered
partially and there is no particular order among the neighborhood. To improve concurrency,
the programmer is encouraged to write rules that take advantage of non-deterministic execution.

LM programs can then be made parallel by simply processing many nodes simultaneously.
Our implementation partitions the graph of N nodes into P subgraphs and then each processing unit will work on its subgraph.
For improved load balancing we use node stealing during starvation.
Our results show that LM programs running on multicores have good scalability.
The implementation of the compiler and virtual machine and
the analysis of experimental results will be presented in a future paper.

%% file: related_work.tex
To the best of our knowledge, LM is the first bottom-up linear logic programming language
that is intended to be executed over graph structures. Although there are a few logic programming languages such as P2~\cite{Loo-condie-garofalakis-p2},
Meld~\cite{ashley-rollman-iclp09}, or Dedalus~\cite{Alvaro:EECS-2009-173} that already do this, they are based on classical logic, where facts are persistent. For most of these systems, there is no concept of state, except for Dedalus where state is modeled as time.

Linear logic has been used in the past as a basis for logic-based programming languages~\cite{Miller85anoverview}, including bottom-up and top-down programming languages. Lolli, a programming language presented in~\cite{Hodas94logicprogramming}, is based on a fragment of intuitionistic linear logic
and proves goals by lazily managing the context of linear resources during top-down proof search. LolliMon~\cite{Lopez:2005:MCL:1069774.1069778} is a concurrent linear logic programming language that integrates both bottom-up and top-down search, where top-down search is done sequentially but bottom-up computations, which are encapsulated inside a monad, can be performed concurrently. Programs start by performing top-down search but this can be suspended in order to perform bottom-up search. This concurrent bottom-up search stops until a fix-point is achieved, after which top-down search is resumed. LolliMon is derived from the concurrent logical framework called CLF~\cite{Watkins:2004uq,Cervesato02aconcurrent,Watkins03aconcurrent}.

Since LM is a bottom-up linear logic programming language, it also shares similarities with Constraint Handling Rules (CHR)~\cite{Betz:2005kx,Betz:2013:LBA:2422085.2422086}.
CHR is a concurrent committed-choice constraint language used to write constraint solvers. A CHR program is a set of rules and
a set of constraints. Constraints can be consumed or generated during the application of rules.
Unlike LM, in CHR there is no
concept of rule priorities, but there is an extension to CHR that supports them~\cite{DeKoninck:2007:URP:1273920.1273924}.
Finally, there is also a CHR extension that adds persistent constraints and it has been proven to be sound and complete~\cite{DBLP:journals/corr/abs-1007-3829}.

Graph Transformation Systems (GTS)~\cite{Ehrig:2004vn}, commonly used to model distributed systems, perform rewriting of graphs through
a set of graph productions. GTS also introduces
concepts of concurrency, where it may be possible to apply several transformations at the same time. In principle, it should be possible to model
LM programs as a graph transformation: we directly map the LM graph of nodes to GTS's initial graph and consider logical facts as nodes that are connected
to LM's nodes. Each LM rule is then a graph production that manipulates the node's neighbors (the database) or sends new facts to other nodes.
On the other hand, it is also possible to embed GTS inside CHR~\cite{Raiser:2011:AGT:1972935.1972938}.

%% file: conclusion.tex
In this paper, we have presented LM, a new linear logic programming language designed with concurrency in mind. LM is a bottom-up logic language
that can naturally model state due to its foundations on linear logic.
We presented several LM programs that show the viability of linear logic programming to solve interesting graph-based problems.

We also gave an overview of the formal system behind LM, namely, the fragment of linear logic used in the language, along with the high level and low level dynamic semantics.
While the former is closely tied to linear logic, the latter is closer to a real implementation.
The low level semantics can be used as a blueprint for someone that intends to implement LM.

%% file: ack.tex
This work is partially funded by the ERDF (European Regional
      Development Fund) through the COMPETE Programme; by FCT (Portuguese
         Foundation for Science and Technology) through the Carnegie Mellon
      Portugal Program and within projects SIBILA
      (NORTE-07-0124-FEDER-000059) and PEst
      (FCOMP-01-0124-FEDER-037281); and by the Qatar National Research Fund under grant NPRP 09-667-1-100.
Flavio Cruz is funded by the FCT grant SFRH / BD / 51566 / 2011.

%% file: appendix.tex
\section{N-Queens program}\label{code:nqueens}

\begin{figure}[h!]
\scriptsize\begin{Verbatim}[numbers=left]
type left(node, node).
type right(node, node).
type down-right(node, node).
type down-left(node, node).
type coord(node, int, int).
type linear propagate-left(node, list node, list int).
type linear propagate-right(node, list node, list int).
type linear test-and-send-down(node, list node, list int).
type linear test-y(node, int, list int, list node, list int).
type linear test-diag-left(node, int, int, list int, list node, list int).
type linear test-diag-right(node, int, int, list int, list node, list int).
type linear send-down(node, list node, list int).
type linear final-state(node, list node, list int).

propagate-right(@0, [], []).

propagate-left(A, Nodes, Coords)
   -o {L | !left(A, L), L <> A | propagate-left(L, Nodes, Coords)}, test-and-send-down(A, Nodes, Coords).
propagate-right(A, Nodes, Coords)
   -o {R | !right(A, R), R <> A | propagate-right(R, Nodes, Coords)}, test-and-send-down(A, Nodes, Coords).

test-and-send-down(A, Nodes, Coords), !coord(A, X, Y)
   -o test-y(A, Y, Coords, Nodes, Coords).

// test if we have a queen on this column
test-y(A, Y, [], Nodes, Coords), !coord(A, OX, OY)
   -o test-diag-left(A, OX - 1, OY - 1, Coords, Nodes, Coords).
test-y(A, Y, [X, Y1 | RestCoords], Nodes, Coords), Y = Y1
   -o 1. // fail
test-y(A, Y, [X, Y1 | RestCoords], Nodes, Coords), Y <> Y1 -o
   test-y(A, Y, RestCoords, Nodes, Coords).

// test if we have a queen on the left diagonal
test-diag-left(A, X, Y, _, Nodes, Coords), X < 0 || Y < 0, !coord(A, OX, OY)
   -o test-diag-right(A, OX - 1, OY + 1, Coords, Nodes, Coords).
test-diag-left(A, X, Y, [X1, Y1 | RestCoords], Nodes, Coords), X = X1, Y = Y1
   -o 1. // fail
test-diag-left(A, X, Y, [X1, Y1 | RestCoords], Nodes, Coords), X <> X1 || Y <> Y1
   -o test-diag-left(A, X - 1, Y - 1, RestCoords, Nodes, Coords).

// test if we have a queen on the right diagonal
test-diag-right(A, X, Y, [], Nodes, Coords), X < 0 || Y >= size, !coord(A, OX, OY)
   -o send-down(A, [A | Nodes], [OX, OY | Coords]). // add new queen
test-diag-right(A, X, Y, [X1, Y1 | RestCoords], Nodes, Coords), X = X1, Y = Y1
   -o 1. // fail
test-diag-right(A, X, Y, [X1, Y1 | RestCoords], Nodes, Coords), X <> X1 || Y <> Y1
   -o test-diag-right(A, X - 1, Y + 1, RestCoords, Nodes, Coords).

send-down(A, Nodes, Coords), !coord(A, size - 1, _)
   -o final-state(A, Nodes, Coords).

send-down(A, Nodes, Coords)
   -o {B | !down-right(A, B), B <> A | propagate-right(B, Nodes, Coords)},
      {B | !down-left(A, B), B <> A | propagate-left(B, Nodes, Coords)}.
\end{Verbatim}
\end{figure}
\normalsize

\clearpage
\section{Linear Logic fragment used in LM}\label{linear_logic}

\begin{figure}[h]
\[
\infer[\one R]
{\Psi ; \seqnocut{\Gamma}{\cdot}{\one}}
{}
\tab
\infer[\one L]
{\Psi ; \seqnocut{\Gamma}{\Delta, \one}{C}}
{\Psi ; \seqnocut{\Gamma}{\Delta}{C}}
\]

\[
\infer[\with R]
{\Psi ; \seqnocut{\Gamma}{\Delta}{A \with B}}
{\Psi ; \seqnocut{\Gamma}{\Delta}{A} & \seqnocut{\Gamma}{\Delta}{B}}
\tab
\infer[\with L_1]
{\Psi ; \seqnocut{\Gamma}{\Delta, A \with B}{C}}
{\Psi ; \seqnocut{\Gamma}{\Delta, A}{C}}
\tab
\infer[\with L_2]
{\Psi ; \seqnocut{\Gamma}{\Delta, B \with B}{C}}
{\Psi ; \seqnocut{\Gamma}{\Delta, B}{C}}
\]

\[
\infer[\otimes R]
{\Psi ; \seqnocut{\Gamma}{\Delta, \Delta'}{A \otimes B}}
{\Psi ; \seqnocut{\Gamma}{\Delta}{A} & \seqnocut{\Gamma}{\Delta}{B}}
\tab
\infer[\otimes L]
{\Psi ;\seqnocut{\Gamma}{\Delta, A \otimes B}{C}}
{\Psi ; \seqnocut{\Gamma}{\Delta, A, B}{C}}
\]

\[
\infer[\lolli R]
{\Psi ; \seqnocut{\Gamma}{\Delta}{A \lolli B}}
{\Psi ; \seqnocut{\Gamma}{\Delta, A}{B}}
\tab
\infer[\lolli L]
{\seqnocut{\Gamma}{\Delta, \Delta', A \lolli B}{C}}
{\Psi ; \seqnocut{\Gamma}{\Delta}{A} &
   \Psi ; \seqnocut{\Gamma}{\Delta', B}{C}}
\]

\[
\infer[\forall R]
{\Psi ; \seqnocut{\Gamma}{\Delta}{\forall n:\tau. A}}
{\Psi, m:\tau ; \seqnocut{\Gamma}{\Delta}{A\{m/n\}}}
\tab
\infer[\forall L]
{\Psi ; \seqnocut{\Gamma}{\Delta, \forall n:\tau. A}{C}}
{\Psi \vdash M : \tau & \Psi ; \seqnocut{\Gamma}{\Delta, A\{M/n\}}{C}}
\]

\[
\infer[\exists R]
{\Psi ; \seqnocut{\Gamma}{\Delta}{\exists n: \tau. A}}
{\Psi \vdash M : \tau &
   \Psi ; \seqnocut{\Gamma}{\Delta}{A \{M/n\}}}
\tab
\infer[\exists L]
{\Psi ; \seqnocut{\Gamma}{\Delta, \exists n:\tau. A}{C}}
{\Psi, m:\tau ; \seqnocut{\Gamma}{\Delta, A\{m/n\}}{C}}
\]

\[
\infer[\bang R]
{\Psi ; \seqnocut{\Gamma}{\cdot}{\bang A}}
{\Psi ; \seqnocut{\Gamma}{\cdot}{A}}
\tab
\infer[\bang L]
{\Psi ; \seqnocut{\Gamma}{\Delta, \bang A}{C}}
{\Psi ; \seqnocut{\Gamma, A}{\Delta}{C}}
\tab
\infer[\m{copy}]
{\Psi ; \seqnocut{\Gamma, A}{\Delta}{C}}
{\Psi ; \seqnocut{\Gamma, A}{\Delta, A}{C}}
\]

\[
\infer[\m{def} \; R]
{\Psi ; \seqnocut{\Gamma}{\Delta}{\compr{A'}}}
{\Psi ; \seqnocut{\Gamma}{\Delta}{B\theta} &
 A \defeq B & A' \doteq A\theta}
\tab
\infer[\m{def} \; L]
{\Psi ; \seqnocut{\Gamma}{\Delta, \compr{A'}}{C}}
{
   \Psi ; \seqnocut{\Gamma}{\Delta, B\theta}{C} & A \defeq B & A' \doteq A\theta
}
\]
\end{figure}

\clearpage
\section{Low Level Dynamic Semantics}\label{low_level_semantics}

All the judgments in this system share a few arguments, namely:

\begin{itemize}
   \item[$\Gamma$]: Set of persistent facts.
   \item[$\Xi'$]: Multi-set of facts consumed after applying one rule (output).
   \item[$\Delta'$]: Multi-set of linear facts derived after applying one rule (output).
   \item[$\Gamma'$]: Set of persistent facts derived after applying one rule (output).
\end{itemize}

\subsection{Application}

The whole process is started by the $\doo$and $\ao$judgments. The $\doo \Gamma; \Delta; \Phi \rightarrow \Xi'; \Delta'; \Gamma'$ judgment starts with facts $\Delta$ and $\Gamma$ and an ordered list of rules that can be applied $\Phi$. The $\ao \Gamma; \Delta; A \lolli B; R \rightarrow \Xi'; \Delta'; \Gamma'$ tries to apply the rule $A \lolli B$ and stores the rule continuation $R$ so that if the current rule fails, we can try another one (in order).

\[
\infer[\ao start \; matching]
{\ao \Gamma; \Delta; A \lolli B; R \rightarrow \Xi'; \Delta'; \Gamma'}
{\mo \Gamma; \Delta; \cdot; A; B; \cdot; R \rightarrow \Xi'; \Delta'; \Gamma'}
\]

\[
\infer[\doo rule]
{\doo \Gamma; \Delta; R, \Phi \rightarrow \Xi'; \Delta';\Gamma'}
{\ao \Gamma; \Delta; R; (\Phi, \Delta) \rightarrow \Xi';\Delta';\Gamma'}
\]

\subsection{Continuation Frames}

Continuation frames are used for backtracking in the matching process.

\subsubsection{Persistent Frame}

A persistent frame has the form $[\Gamma'; \Delta; \Xi; \bang p; \Omega; \Lambda; \Upsilon]$, where:

\begin{enumerate}
   \item[$\Delta$]: Remaining multi-set of linear facts.
   \item[$\Xi$]: Multi-set of linear facts we have consumed to reach this point.
   \item[$\bang p$]: Current fact expression that originated this choice point.
   \item[$\Omega$]: Remaining terms we need to match past this choice point. This is an ordered list.
   \item[$\Lambda$]: Multi-set of linear fact expressions that we have matched to reach this choice point. All the linear facts that match these terms are located in $\Xi$.
   \item[$\Upsilon$]: Multi-set of persistent fact expressions that we matched up to this point.
\end{enumerate}

\subsubsection{Linear Frame}

A linear frame has the form $(\Delta; \Delta'; \Xi; p; \Omega; \Lambda; \Upsilon)$, where:

\begin{enumerate}
   \item[$\Delta$]: Multi-set of linear facts that are not of type $p$ plus all the other $p$'s we have already tried, including the current $p$.
   \item[$\Delta'$]: All the other $p$'s we haven't tried yet. It is a multi-set of linear facts.
   \item[$\Xi$]: Multi-set of linear facts we have consumed to reach this point.
   \item[$p$]: Current fact expression that originated this choice point.
   \item[$\Omega$]: Remaining terms we need to match past this choice point. This is an ordered list.
   \item[$\Lambda$]: Multi-set of linear fact expressions that we have matched to reach this choice point. All the linear facts that match these terms are located in $\Xi$.
   \item[$\Upsilon$]: Multi-set of persistent fact expressions that we matched up to this point.
\end{enumerate}

\subsection{Match}

The matching judgment uses the form $\mo \Gamma; \Delta; \Xi; \Omega; H; C; R \rightarrow \Xi'; \Delta'; \Gamma'$ where:

\begin{enumerate}
   \item[$\Delta$]: Multi-set of linear facts still available to complete the matching process.
   \item[$\Xi$]: Multi-set of linear facts consumed up to this point.
   \item[$\Omega$]: Ordered list of terms we want to match.
   \item[$H$]: Head of the rule.
   \item[$C$]: Ordered list of frames representing the continuation stack.
   \item[$R$]: Rule continuation. If the matching process fails, we try another rule.
\end{enumerate}

Matching will attempt to use facts from $\Delta$ and $\Gamma$ to match the terms of the body of the rule represented as $\Omega$. During this process continuation frames are pushed into $C$.
If the matching process fails, we use the continuation stack through the $\cont$judgment.

{\footnotesize
\[
\infer[\mo p \; ok \; first]
{\mo \Gamma ; \Delta, p_1, \Delta'' ; \Xi; p, \Omega; H; \cdot; R \rightarrow \Xi'; \Delta'; \Gamma'}
{\mo \Gamma ; \Delta, \Delta''; \Xi, p_1; \Omega; H; (\Delta, p_1; \Delta''; p; \Omega; \Xi; \cdot; \cdot); R \rightarrow \Xi'; \Delta'; \Gamma'}
\]

\[
\infer[\mo p \; ok \; other \; q]
{\mo \Gamma ; \Delta, p_1, \Delta'' ; \Xi; p, \Omega; H; C_1, C; R \rightarrow \Xi'; \Delta'; \Gamma'}
{\begin{split}\mo &\Gamma ; \Delta, \Delta''; \Xi, p_1; \Omega; H; (\Delta, p_1; \Delta''; p; \Omega; \Xi; q, \Lambda; \Upsilon), C_1, C; R \rightarrow \Xi'; \Delta'; \Gamma' \\ C_1 &= (\Delta_{old}; \Delta'_{old}; q; \Omega_{old}; \Xi_{old}; \Lambda; \Upsilon)\end{split}}
\]

\[
\infer[\mo p \; ok \; other \; \bang q]
{\mo \Gamma ; \Delta, p_1, \Delta'' ; \Xi; p, \Omega; H; C_1, C; R \rightarrow \Xi'; \Delta'; \Gamma'}
{\begin{split}\mo &\Gamma ; \Delta, \Delta''; \Xi, p_1; \Omega; H; (\Delta, p_1; \Delta''; p; \Omega; \Xi; \Lambda; q, \Upsilon), C_1, C; R \rightarrow \Xi'; \Delta'; \Gamma' \\ C_1 &= [\Gamma_{old}; \Delta_{old}; \bang q; \Omega_{old}; \Xi_{old}; \Lambda; \Upsilon]\end{split}}
\]

\[
\infer[\mo p \; fail]
{\mo \Gamma ; \Delta; \Xi; p, \Omega; H; C; R \rightarrow \Xi'; \Delta'; \Gamma'}
{p \notin \Delta & \cont C ; H; R; \Gamma; \Xi'; \Delta'; \Gamma'}
\]

\[
\infer[\mo \bang p \; ok \; first]
{\mo \Gamma, p_1, \Gamma'' ; \Delta; \Xi; \bang p, \Omega; H; \cdot; R \rightarrow \Xi'; \Delta'; \Gamma'}
{\mo \Gamma, p_1, \Gamma'' ; \Delta; \Xi; \Omega; H; [\Gamma''; \Delta; \bang p ; \Omega; \Xi; \cdot; \cdot]; R \rightarrow \Xi'; \Delta'; \Gamma'}
\]

\[
\infer[\mo \bang p \; ok \; other \; q]
{\mo \Gamma, p_1, \Gamma'' ; \Delta; \Xi; \bang p, \Omega; H; C_1, C; R \rightarrow \Xi'; \Delta'; \Gamma'}
{\begin{split}\mo &\Gamma, p_1, \Gamma'' ; \Delta; \Xi; \Omega; H; [\Gamma''; \Delta; \bang p ; \Omega; \Xi; q, \Lambda; \Upsilon], C_1, C; R \rightarrow \Xi'; \Delta'; \Gamma' \\ C_1 &= (\Delta_{old}; \Delta'_{old}; q; \Omega_{old}; \Xi_{old}; \Lambda; \Upsilon)\end{split}}
\]

\[
\infer[\mo \bang p \; ok \; other \; \bang q]
{\mo \Gamma, p_1, \Gamma'' ; \Delta; \Xi; \bang p, \Omega; H; C_1, C; R \rightarrow \Xi'; \Delta'; \Gamma'}
{\begin{split}\mo &\Gamma, p_1, \Gamma'' ; \Delta; \Xi; \Omega; H; [\Gamma''; \Delta; \bang p ; \Omega; \Xi; \Lambda; q, \Upsilon], C_1, C; R \rightarrow \Xi'; \Delta'; \Gamma' \\ C_1 &= [\Gamma_{old}; \Delta_{old}; \bang q; \Omega_{old}; \Xi_{old}; \Lambda; \Upsilon]\end{split}}
\]

\[
\infer[\mo \bang p \; fail]
{\mo \Gamma ; \Delta; \Xi; \bang p, \Omega; H; C; R \rightarrow \Xi'; \Delta'; \Gamma'}
{\bang p \notin \Gamma & \cont C; H; R; \Gamma; \Xi'; \Delta'; \Gamma'}
\]

\[
\infer[\mo \otimes]
{\mo \Gamma ; \Delta; \Xi; A \otimes B, \Omega ; H ; C; R \rightarrow \Xi'; \Delta';\Gamma'}
{\mo \Gamma ; \Delta; \Xi; A, B, \Omega; H; C; R \rightarrow \Xi';\Delta';\Gamma'}
\]

\[
\infer[\mo end]
{\mo \Gamma ; \Delta; \Xi; \cdot ; H; C; R \rightarrow \Xi'; \Delta'; \Gamma'}
{\done \Gamma ; \Delta; \Xi; \cdot ; H; \cdot \rightarrow \Xi'; \Delta'; \Gamma'}
\]
}

\subsection{Continuation}

If the previous matching process fails, we pick the top frame from the stack $C$ and restore the matching process using another fact and/or context. The continuation judgment $\cont C; H; R; \Gamma; \Xi'; \Delta'; \Gamma'$ deals with the backtracking process where the meaning of each argument is as follows:

\begin{enumerate}
   \item[$C$]: Continuation stack.
   \item[$H$]: Head of the current rule we are trying to match.
   \item[$R$]: Next available rules if the current one does not match.
\end{enumerate}

{\footnotesize
\[
\infer[\cont next \; rule]
{\cont \cdot; H; (\Phi, \Delta); \Gamma ; \Xi'; \Delta'; \Gamma'}
{\doo \Gamma; \Delta; \Phi \rightarrow \Xi'; \Delta'; \Gamma'}
\]

\[
\infer[\cont p \; next]
{\cont (\Delta; p_1, \Delta''; p, \Omega; \Xi; \Lambda; \Upsilon), C; H; R; \Gamma; \Xi'; \Delta'; \Gamma'}
{\mo \Gamma ; \Delta, \Delta''; \Xi, p_1; \Omega; H; (\Delta, p_1; \Delta''; p, \Omega; H; \Xi; \Lambda; \Upsilon), C; R \rightarrow \Xi'; \Delta'; \Gamma'}
\]

\[
\infer[\cont p \; no \; more]
{\cont (\Delta; \cdot; p, \Omega; \Xi; \Lambda; \Upsilon), C; H; R; \Gamma; \Xi'; \Delta'; \Gamma'}
{\cont C; H; R; \Gamma; \Xi'; \Delta'; \Gamma'}
\]

\[
\infer[\cont \bang p \; next]
{\cont [p_1, \Gamma'; \Delta; \bang p, \Omega; \Xi; \Lambda; \Upsilon], C; H; R; \Gamma; \Xi'; \Delta'; \Gamma'}
{\mo \Gamma; \Delta; \Xi; \Omega; H; [\Gamma'; \Delta; \bang p, \Omega; \Xi; \Lambda; \Upsilon], C; R \rightarrow \Xi'; \Delta'; \Gamma'}
\]

\[
\infer[\cont \bang p \; no \; more]
{\cont [\cdot; \Delta; \bang p, \Omega; \Xi; \Lambda; \Upsilon], C; H; R; \Gamma; \Xi'; \Delta'; \Gamma'}
{\cont C; H; R; \Gamma; \Xi'; \Delta'; \Gamma'}
\]
}

\subsection{Derivation}

Once the list of terms $\Omega$ in the $\mo$judgment is exhausted, we derive the terms of the head of rule.
The derivation judgment uses the form $\done \Gamma; \Delta; \Xi; \Gamma_1; \Delta_1; \Omega \rightarrow \Xi'; \Delta'; \Gamma'$:

\begin{enumerate}
   \item[$\Delta$]: Multi-set of linear facts we started with minus the linear facts consumed during the matching of the body of the rule.
   \item[$\Xi$]: Multi-set of linear facts consumed during the matching of the body of the rule.
   \item[$\Gamma_1$]: Set of persistent facts derived up to this point in the derivation.
   \item[$\Delta_1$]: Multi-set of linear facts derived up to this point in the derivation.
   \item[$\Omega$]: Remaining terms to derive as an ordered list. We start with $B$ if the original rule is $A \lolli B$.
\end{enumerate}

{\footnotesize
\[
\infer[\done p]
{\done \Gamma ; \Delta; \Xi; \Gamma_1 ; \Delta_1; p, \Omega \rightarrow \Xi'; \Delta'; \Gamma'}
{\done \Gamma ; \Delta; \Xi; \Gamma_1 ; p, \Delta_1; \Omega \rightarrow \Xi'; \Delta'; \Gamma'}
\tab
\infer[\done 1]
{\done \Gamma; \Delta; \Xi; \Gamma_1 ; \Delta_1; 1, \Omega \rightarrow \Xi';\Delta';\Gamma'}
{\done \Gamma; \Delta; \Xi; \Gamma_1 ; \Delta_1; \Omega \rightarrow \Xi'; \Delta';\Gamma'}
\]

\[
\infer[\done \bang p]
{\done \Gamma ; \Delta ; \Xi; \Gamma_1 ; \Delta_1; \bang p, \Omega \rightarrow \Xi'; \Delta'; \Gamma'}
{\done \Gamma ; \Delta ; \Xi; \Gamma_1, p; \Delta_1; \Omega \rightarrow \Xi'; \Delta'; \Gamma'}
\]

\[
\infer[\done \otimes]
{\done \Gamma ; \Delta; \Xi; \Gamma_1; \Delta_1; A \otimes B, \Omega \rightarrow \Xi'; \Delta';\Gamma'}
{\done \Gamma ; \Delta; \Xi; \Gamma_1; \Delta_1; A, B, \Omega \rightarrow \Xi';\Delta';\Gamma'}
\]

\[
\infer[\done end]
{\done \Gamma; \Delta; \Xi; \Gamma_1; \Delta_1; \cdot \rightarrow \Xi; \Delta_1; \Gamma_1}
{}
\]

\[
\infer[\done comp]
{\done \Gamma; \Delta ; \Xi; \Gamma_1; \Delta_1; \comp A \lolli B, \Omega \rightarrow \Xi';\Delta';\Gamma'}
{\mc \Gamma; \Delta; \Xi; \Gamma_1; \Delta_1; \cdot; A ; B ; \cdot; \cdot; \comp A \lolli B; \Omega; \Delta \rightarrow \Xi';\Delta';\Gamma'}
\]
}

\subsection{Match Comprehension}

The matching process for comprehensions is similar to the process used for matching the body of the rule, however
we use two continuation stacks, $C$ and $P$. In $P$, we put all the initial persistent frames and in $C$ we put the first linear frame and then everything else.
With this we can easily find out the first linear frame and remove everything that was pushed on top of such frame.
The full judgment has the form
$\mc \Gamma; \Delta; \Xi_N; \Gamma_{N1}; \Delta_{N1}; \Xi; \Omega; C; P; AB; \Omega_N; \Delta_N \rightarrow \Xi'; \Delta'; \Gamma'$:

\begin{enumerate}
   \item[$\Delta$]: Multi-set of linear facts remaining up to this point in the matching process.
   \item[$\Xi_N$]: Multi-set of linear facts used during the matching process of the body of the rule.
   \item[$\Gamma_{N1}$]: Set of persistent facts derived up to this point in the head of the rule and all the previous comprehensions.
   \item[$\Delta_{N1}$]: Multi-set of linear facts derived by the head of the rule and by the previous comprehensions.
   \item[$\Xi$]: Multi-set of linear facts consumed up to this point.
   \item[$\Omega$]: Ordered list of terms that need to be matched for the comprehension to be applied.
   \item[$C$]: Continuation stack that contains both linear and persistent frames. The first frame must be linear.
   \item[$P$]: Initial part of the continuation stack with only persistent frames.
   \item[$AB$]: Comprehension $\comp A \lolli B$ that is being matched.
   \item[$\Omega_N$]: Ordered list of remaining terms of the head of the rule to be derived.
   \item[$\Delta_N$]: Multi-set of linear facts that were still available after matching the body of the rule and the previous comprehensions. Note that $\Delta, \Xi = \Delta_N$.
\end{enumerate}

{\scriptsize

\[
\infer[\mc p \; ok \; first]
{\mc \Gamma; \Delta, p_1, \Delta''; \Xi_N; \Gamma_{N1}; \Delta_{N1}; \cdot; p, \Omega; \cdot; \cdot; AB; \Omega_N; \Delta_N \rightarrow \Xi'; \Delta'; \Gamma'}
{\mc \Gamma; \Delta, \Delta''; \Xi_N; \Gamma_{N1}; \Delta_{N1}; \Xi, p_1; \Omega; (\Delta, p_1; \Delta''; \cdot; p; \Omega; \cdot; \cdot); \cdot; AB; \Omega_N; \Delta_N \rightarrow \Xi'; \Delta'; \Gamma'}
\]

\[
\infer[\mc p \; ok \; other \; q]
{\mc \Gamma; \Delta, p_1, \Delta''; \Xi_N; \Gamma_{N1}; \Delta_{N1}; \Xi; p, \Omega; C_1, C; P; AB; \Omega_N; \Delta_N \rightarrow \Xi'; \Delta'; \Gamma'}
{\begin{split}\mc &\Gamma; \Delta, \Delta''; \Xi_N; \Gamma_{N1}; \Delta_{N1}; \Xi, p_1; \Omega; (\Delta, p_1; \Delta''; \Xi; p; \Omega; q, \Lambda; \Upsilon), C_1, C; P; AB; \Omega_N; \Delta_N \rightarrow \Xi'; \Delta'; \Gamma' \\ C_1 &= (\Delta_{old}; \Delta'_{old}; \Xi_{old}; q; \Omega_{old}; \Lambda; \Upsilon)\end{split}}
\]

\[
\infer[\mc p \; ok \; other \; \bang qC]
{\mc \Gamma; \Delta, p_1, \Delta''; \Xi_N; \Gamma_{N1}; \Delta_{N1}; \Xi; p, \Omega; C_1, C; P; AB; \Omega_N; \Delta_N \rightarrow \Xi'; \Delta'; \Gamma'}
{\begin{split}\mc &\Gamma; \Delta, \Delta''; \Xi_N; \Gamma_{N1}; \Delta_{N1}; \Xi, p_1; \Omega; (\Delta, p_1; \Delta''; \Xi; p; \Omega; \Lambda; q, \Upsilon), C_1, C; P; AB; \Omega_N; \Delta_N \rightarrow \Xi'; \Delta'; \Gamma' \\ C_1 &= [\Gamma_{old}; \Delta_{old}; \Xi_{old}; q; \Omega_{old}; \Lambda; \Upsilon]\end{split}}
\]

\[
\infer[\mc p \; ok \; other \; \bang qP]
{\mc \Gamma; \Delta, p_1, \Delta''; \Xi_N; \Gamma_{N1}; \Delta_{N1}; \cdot; p, \Omega; \cdot; P_1, P; AB; \Omega_N; \Delta_N \rightarrow \Xi'; \Delta'; \Gamma'}
{\begin{split}\mc &\Gamma; \Delta, \Delta''; \Xi_N; \Gamma_{N1}; \Delta_{N1}; p_1; \Omega; (\Delta, p_1; \Delta''; \cdot; p; \Omega; \cdot; q, \Upsilon); P_1, P; AB; \Omega_N; \Delta_N \rightarrow \Xi'; \Delta'; \Gamma' \\ P_1 &= [\Gamma_{old}; \Delta_N; \cdot; q; \Omega_{old}; \cdot; \Upsilon]\\ \Delta_N &= \Delta, p_1, \Delta''\end{split}}
\]

\[
\infer[\mc p \; fail]
{\mc \Gamma; \Delta; \Xi_N; \Gamma_{N1}; \Delta_{N1}; \Xi; p, \Omega; C; P; AB; \Omega_N; \Delta_N \rightarrow \Xi'; \Delta'; \Gamma'}
{p \notin \Delta & \contc \Gamma; \Delta_N; \Xi_N; \Gamma_{N1}; \Delta_{N1}; C; P; AB; \Omega_N \rightarrow \Xi'; \Delta'; \Gamma'}
\]

\[
\infer[\mc \bang p \; first]
{\mc \Gamma, \Gamma'', p; \Delta_N; \Xi_N; \Gamma_{N1}; \Delta_{N1}; \cdot; \bang p, \Omega; \cdot; \cdot; AB; \Omega_N; \Delta_N \rightarrow \Xi'; \Delta'; \Gamma'}
{\mc \Gamma, \Gamma'', p; \Delta_N; \Xi_N; \Gamma_{N1}; \Delta_{N1}; \cdot; \Omega; \cdot; [\Gamma''; \Delta_N; \cdot; \bang p; \cdot; \Omega; \cdot; \cdot]; AB; \Omega_N; \Delta_N \rightarrow \Xi'; \Delta'; \Gamma'}
\]

\[
\infer[\mc \bang p \; other \; \bang qP]
{\mc \Gamma, \Gamma'', p; \Delta_N; \Xi_N; \Gamma_{N1}; \Delta_{N1}; \cdot; \bang p, \Omega; \cdot; P_1, P; AB; \Omega_N; \Delta_N \rightarrow \Xi'; \Delta'; \Gamma'}
{\begin{split}\mc &\Gamma, \Gamma'', p; \Delta_N; \Xi_N; \Gamma_{N1}; \Delta_{N1}; \cdot; \Omega; [\Gamma''; \Delta_N; \cdot; \bang p; \cdot; \Omega; \cdot; q, \Upsilon], P_1, P; AB; \Omega_N; \Delta_N \rightarrow \Xi'; \Delta'; \Gamma' \\ P_1 &= [\Gamma_{old}; \Delta_N; \cdot; \bang q; \Omega_{old}; \cdot; \Upsilon]\end{split}}
\]

\[
\infer[\mc \bang p \; other \; \bang qC]
{\mc \Gamma, \Gamma'', p; \Delta; \Xi_N; \Gamma_{N1}; \Delta_{N1}; \Xi; \bang p, \Omega; C_1, C; P; AB; \Omega_N; \Delta_N \rightarrow \Xi'; \Delta'; \Gamma'}
{\begin{split}\mc &\Gamma, \Gamma'', p; \Delta; \Xi_N; \Gamma_{N1}; \Delta_{N1}; \Xi; \Omega; [\Gamma''; \Delta; \Xi; \bang p; \cdot; \Omega; \Lambda; q, \Upsilon], C_1, C; P; AB; \Omega_N; \Delta_N \rightarrow \Xi'; \Delta'; \Gamma' \\ C_1 &= [\Gamma_{old}; \Delta_{old}; \Xi_{old}; \bang q; \Omega_{old}; \Lambda; \Upsilon]\end{split}}
\]

\[
\infer[\mc \bang p \; other \; qC]
{\mc \Gamma, \Gamma'', p; \Delta; \Xi_N; \Gamma_{N1}; \Delta_{N1}; \Xi; \bang p, \Omega; C_1, C; P; AB; \Omega_N; \Delta_N \rightarrow \Xi'; \Delta'; \Gamma'}
{\begin{split}\mc &\Gamma, \Gamma'', p; \Delta; \Xi_N; \Gamma_{N1}; \Delta_{N1}; \Xi; \Omega; [\Gamma''; \Delta; \Xi; \bang p; \cdot; \Omega; \Lambda, q; \Upsilon], C_1, C; P; AB; \Omega_N; \Delta_N \rightarrow \Xi'; \Delta'; \Gamma' \\ C_1 &= (\Delta_{old}; \Delta'_{old}; \Xi_{old}; q; \Omega_{old}; \Lambda; \Upsilon)\end{split}}
\]

\[
\infer[\mc \bang p~fail]
{\mc \Gamma; \Delta; \Xi_N; \Gamma_{N1}; \Delta_{N1}; \Xi; \bang p, \Omega; C; P; AB; \Omega_N; \Delta_N \rightarrow \Xi'; \Delta; \Gamma'}
{\bang p \notin \Gamma & \contc \Gamma; \Delta_N; \Xi_N; \Gamma_{N1}; \Delta_{N1}; C; P; AB; \Omega_N \rightarrow \Xi'; \Delta'; \Gamma'}
\]

\[
\infer[\mc \otimes]
{\mc \Delta; \Xi_N; \Delta_{N1}; \Xi; X \otimes Y, \Omega; C; P; AB; \Omega_N; \Delta_N \rightarrow \Xi'; \Delta'}
{\mc \Delta; \Xi_N; \Delta_{N1}; \Xi; X, Y, \Omega; C; P; AB; \Omega_N; \Delta_N \rightarrow \Xi'; \Delta'}
\]

\[
\infer[\mc 1]
{\mc \Delta; \Xi_N; \Delta_{N1}; \Xi; 1, \Omega; C; P; AB; \Omega_N; \Delta_N \rightarrow \Xi'; \Delta'}
{\mc \Delta; \Xi_N; \Delta_{N1}; \Xi; \Omega; C; P; AB; \Omega_N; \Delta_N \rightarrow \Xi'; \Delta'}
\]

\[
\infer[\mc end]
{\mc \Gamma; \Delta; \Xi_N; \Gamma_{N1}; \Delta_{N1}; \Xi; \cdot; C; P; AB; \Omega_N; \Delta_N \rightarrow \Xi'; \Delta'; \Gamma'}
{\dall \Gamma; \Xi_N; \Gamma_{N1}; \Delta_{N1}; \Xi; C; P; AB; \Omega_N; \Delta_N \rightarrow \Xi'; \Delta'; \Gamma'}
\]
}

\subsection{Match Comprehension Continuation}

If the matching process fails, we need to backtrack to the previous frame and restore the matching process at that point. The judgment that backtracks has the form $\contc \Gamma; \Delta_N; \Xi_N; \Delta_{N1}; C; P; AB; \Omega_N \rightarrow \Xi'; \Delta'; \Gamma'$:

\begin{enumerate}
   \item[$\Delta_N$]: Multi-set of linear facts that were still available after matching the body of the rule.
   \item[$\Xi_N$]: Multi-set of linear facts used to match the body of the rule and all the previous comprehensions.
   \item[$\Gamma_{N1}$]: Set of persistent facts derived by the head of the rule and all the previous comprehensions.
   \item[$\Delta_{N1}$]: Multi-set of linear facts derived by the head of the rule and all the previous comprehensions.
   \item[$C, P$]: Continuation stacks.
   \item[$AB$]: Comprehension $\comp A \lolli B$ that is being matched.
   \item[$\Omega_N$]: Ordered list of remaining terms of the head of the rule to be derived.
\end{enumerate}

{\small
\[
\infer[\contc end]
{\contc \Gamma; \Delta_N; \Xi_N; \Gamma_{N1}; \Delta_{N1}; \cdot; \cdot; \comp A \lolli B; \Omega \rightarrow \Xi'; \Delta'; \Gamma'}
{\done \Gamma; \Delta_N; \Xi_N; \Gamma_{N1}; \Delta_{N1}; \Omega \rightarrow \Xi'; \Delta'; \Gamma'}
\]

\[
\infer[\contc nextC \; p]
{\contc \Gamma; \Delta_N; \Xi_N; \Gamma_{N1}; \Delta_{N1}; (\Delta; p_1, \Delta''; \Xi; p; \Omega; \Lambda; \Upsilon), C; P; AB; \Omega_N \rightarrow \Xi'; \Delta'; \Gamma'}
{\mc \Gamma; \Delta; \Xi_N; \Gamma_{N1}; \Delta_{N1}; \Xi; \Omega; (\Delta, p_1; \Delta''; \Xi; p; \Omega; \Lambda; \Upsilon), C; P; AB; \Omega_N; \Delta_N \rightarrow \Xi'; \Delta'; \Gamma'}
\]

\[
\infer[\contc nextC \; \bang p]
{\contc \Gamma; \Delta_N; \Xi_N; \Gamma_{N1}; \Delta_{N1}; [p_1, \Gamma'; \Delta; \Xi; \bang p; \Omega; \Lambda; \Upsilon], C; P; AB; \Omega_N \rightarrow \Xi'; \Delta'; \Gamma'}
{\mc \Gamma; \Delta; \Xi_N; \Gamma_{N1}; \Delta_{N1}; \Xi; \Omega; [\Gamma'; \Delta; \Xi; \bang p; \Omega; \Lambda; \Upsilon], C; P; AB; \Omega_N; \Delta_N \rightarrow \Xi'; \Delta'; \Gamma'}
\]

\[
\infer[\contc nextC \; empty \; p]
{\contc \Gamma; \Delta_N; \Xi_N; \Gamma_{N1}; \Delta_{N1}; (\Delta; \cdot; \Xi; p; \Omega; \Lambda; \Upsilon), C; P; AB; \Omega_N \rightarrow \Xi'; \Delta'; \Gamma'}
{\contc \Gamma; \Delta_N; \Xi_N; \Gamma_{N1}; \Delta_{N1}; C; P; AB; \Omega_N \rightarrow \Xi'; \Delta'; \Gamma'}
\]

\[
\infer[\contc nextC \; empty \; \bang p]
{\contc \Gamma; \Delta_N; \Xi_N; \Gamma_{N1}; \Delta_{N1}; [\cdot; \Delta; \Xi; \bang p; \Omega; \Lambda; \Upsilon], C; P; AB; \Omega_N \rightarrow \Xi'; \Delta'; \Gamma'}
{\contc \Gamma; \Delta_N; \Xi_N; \Gamma_{N1}; \Delta_{N1}; C; P; AB; \Omega_N \rightarrow \Xi'; \Delta'; \Gamma'}
\]

\[
\infer[\contc nextP \; \bang p]
{\contc \Gamma; \Delta_N; \Xi_N; \Gamma_{N1}; \Delta_{N1}; \cdot; [p_1, \Gamma'; \Delta_N; \cdot; \bang p; \Omega; \cdot; \Upsilon], P; AB; \Omega_N \rightarrow \Xi'; \Delta'; \Gamma'}
{\mc \Gamma; \Delta_N; \Xi_N; \Gamma_{N1}; \Delta_{N1}; \cdot; \Omega; \cdot; [\Gamma'; \Delta_N; \cdot; \bang p; \Omega; \cdot; \Upsilon], P; AB; \Omega_N; \Delta_N \rightarrow \Xi'; \Delta'; \Gamma'}
\]

\[
\infer[\contc nextP \; empty \; \bang p]
{\contc \Gamma; \Delta_N; \Xi_N; \Gamma_{N1}; \Delta_{N1}; \cdot; [\cdot; \Delta_N; \cdot; \bang p; \Omega; \cdot; \Upsilon], P; AB; \Omega_N \rightarrow \Xi'; \Delta'; \Gamma'}
{\contc \Gamma; \Delta_N; \Xi_N; \Gamma_{N1}; \Delta_{N1}; \cdot; P; AB; \Omega_N \rightarrow \Xi'; \Delta'; \Gamma'}
\]
}

\subsection{Stack Transformation}

After a comprehension is matched and before we derive the head of such comprehension, we need to "fix" the continuation stack. In a nutshell, we remove all the frames except the first linear frame and remove the consumed linear facts from the remaining frames so that they are still valid for the next application of the comprehension.
The judgment that fixes the stack has the form $\dall \Gamma; \Xi_N; \Gamma_{N1}; \Delta_{N1}; \Xi; C; P; AB; \Omega_N; \Delta_N \rightarrow \Xi'; \Delta'; \Gamma'$:

\begin{enumerate}
   \item[$\Xi_N$]: Multi-set of linear facts used during the matching process of the body of the rule.
   \item[$\Gamma_{N1}$]: Set of persistent facts derived by the head of the rule and all the previous comprehensions.
   \item[$\Delta_{N1}$]: Multi-set of linear facts derived by the head of the rule and all the previous comprehensions.
   \item[$\Xi$]: Multi-set of linear facts consumed by the previous application of the comprehension.
   \item[$C, P$]: Continuation stacks for the comprehension.
   \item[$AB$]: Comprehension $\comp A \lolli B$ that is being matched.
   \item[$\Omega_N$]: Ordered list of remaining terms of the head of the rule to be derived.
   \item[$\Delta_N$]: Multi-set of linear facts that were still available after matching the body of the rule and all the previous comprehensions.
\end{enumerate}

{\footnotesize
\[
\infer[\strans]
{\strans \Xi; [\Gamma'; \Delta_N; \cdot; \bang p; \Omega; \cdot; \Upsilon], P; [\Gamma'; \Delta_N - \Xi; \cdot; \bang p, \Omega; \cdot; \Upsilon], P'}
{\strans \Xi; P; P'}
\]

\[
\infer[\strans end]
{\strans \Xi; \cdot; \cdot}
{\strans \Xi; \cdot; \cdot}
\]

\[
\infer[\dall end \; linear]
{\dall \Gamma; \Xi_N; \Gamma_{N1}; \Delta_{N1}; \Xi; (\Delta_x; \Delta''; \cdot; p; \Omega; \cdot; \Upsilon); P;  \comp A \lolli B; \Omega_N; \Delta_N \rightarrow \Xi'; \Delta'; \Gamma'}
{\begin{split}\strans &\Xi; P; P' \\ \dc \Gamma; \Xi_N, \Xi; \Gamma_{N1}; \Delta_{N1}; (\Delta_x - \Xi; \Delta'' - \Xi; \cdot; p; \Omega; \cdot; \Upsilon) ; P' ; \comp A \lolli B; \Omega_N; (\Delta_N - \Xi) &\rightarrow \Xi'; \Delta'; \Gamma'\end{split}}
\]

\[
\infer[\dall end \; empty]
{\dall \Gamma; \Xi_N; \Gamma_{N1}; \Delta_{N1}; \Xi; \cdot; P;  \comp A \lolli B; \Omega_N; \Delta_N \rightarrow \Xi'; \Delta'; \Gamma'}
{\begin{split}\strans &\Xi; P; P' \\ \dc \Gamma; \Xi_N, \Xi; \Gamma_{N1}; \Delta_{N1}; \cdot ; P' ; \comp A \lolli B; \Omega_N; (\Delta_N - \Xi) &\rightarrow \Xi'; \Delta'; \Gamma'\end{split}}
\]

\[
\infer[\dall more]
{\dall \Gamma; \Xi_N; \Gamma_{N1}; \Delta_{N1}; \Xi; \_, X, C; P; \comp A \lolli B; \Omega_N; \Delta_N \rightarrow \Xi'; \Delta'; \Gamma'}
{\dall \Gamma; \Xi_N; \Gamma_{N1}; \Delta_{N1}; \Xi; X, C; P; \comp A \lolli B; \Omega_N; \Delta_N \rightarrow \Xi'; \Delta'; \Gamma'}
\]
}

\subsection{Comprehension Derivation}

After the fixing process, we start deriving the head of the comprehension. All the facts derived go directly to $\Gamma_{1}$ and $\Delta_{1}$. The judgment has the form $\dc \Gamma; \Xi; \Gamma_1; \Delta_1; \Omega; C; P; AB; \Omega_N; \Delta_N \rightarrow \Xi'; \Delta'; \Gamma'$:

\begin{enumerate}
   \item[$\Xi$]: Multi-set of linear facts consumed both by the body of the rule and all the comprehension applications.
   \item[$\Gamma_1$]: Set of persistent facts derived by the head of the rule and comprehensions.
   \item[$\Delta_1$]: Multi-set of linear facts derived by the head of the rule and comprehensions.
   \item[$\Omega$]: Ordered list of terms to derive.
   \item[$C, P$]: New continuation stacks.
   \item[$AB$]: Comprehension $\comp A \lolli B$ that is being derived.
   \item[$\Omega_N$]: Ordered list of remaining terms of the head of the rule to be derived.
   \item[$\Delta_N$]: Multi-set of remaining linear facts that can be used for the next comprehension applications.
\end{enumerate}

{\small
\[
\infer[\dc p]
{\dc \Gamma; \Xi; \Gamma_1; \Delta_1; p, \Omega; C; P; \comp A \lolli B; \Omega_N; \Delta_N \rightarrow \Xi'; \Delta'; \Gamma'}
{\dc \Gamma; \Xi; \Gamma_1; \Delta_1, p; \Omega; C; P; \comp A \lolli B; \Omega_N; \Delta_N \rightarrow \Xi'; \Delta'; \Gamma'}
\]

\[
\infer[\dc \bang p]
{\dc \Gamma; \Xi; \Gamma_1; \Delta_1; \bang p, \Omega; C; P; \comp A \lolli B; \Omega_N; \Delta_N \rightarrow \Xi'; \Delta'; \Gamma'}
{\dc \Gamma; \Xi; \Gamma_1, p; \Delta_1; \Omega; C; P; \comp A \lolli B; \Omega_N; \Delta_N \rightarrow \Xi'; \Delta'; \Gamma'}
\]

\[
\infer[\dc \otimes]
{\dc \Gamma; \Xi; \Gamma_1; \Delta_1; A \otimes B, \Omega; C; P; \comp A \lolli B; \Omega_N; \Delta_N \rightarrow \Xi'; \Delta';\Gamma'}
{\dc \Gamma; \Xi; \Gamma_1; \Delta_1; A, B, \Omega; C; P; \comp A \lolli B; \Omega_N; \Delta_N \rightarrow \Xi'; \Delta';\Gamma'}
\]

\[
\infer[\dc 1]
{\dc \Gamma; \Xi; \Gamma_1; \Delta_1; 1, \Omega; C; P; \comp A \lolli B; \Omega_N; \Delta_N \rightarrow \Xi'; \Delta';\Gamma'}
{\dc \Gamma; \Xi; \Gamma_1; \Delta_1; \Omega; C; P; \comp A \lolli B; \Omega_N; \Delta_N \rightarrow \Xi'; \Delta';\Gamma'}
\]

\[
\infer[\dc end]
{\dc \Gamma; \Xi; \Gamma_1; \Delta_1; \cdot; C; P; \comp A \lolli B; \Omega_N; \Delta_N \rightarrow \Xi'; \Delta'; \Gamma'}
{\contc \Gamma; \Delta_N; \Xi; \Gamma_1; \Delta_1; C; P; \comp A \lolli B; \Omega_N \rightarrow \Xi'; \Delta'; \Gamma'}
\]
}